\documentclass[11pt,a4paper]{article}
\usepackage[hyperref]{nlp4MusA}
\usepackage{times}
\usepackage{latexsym}
\usepackage{tikz}
\usepackage{adjustbox}
\usepackage{multicol}
\usepackage{graphicx}
\usepackage{subcaption}
\usepackage{color}
\usepackage{xcolor}
\usepackage{arydshln}
\usepackage{soul}

\usepackage{url}

\usepackage{booktabs}

\usepackage{pifont}

\newcommand{\eg}{e.\,g., }
\newcommand{\ie}{i.\,e., }

\aclfinalcopy 

\title{The Role of Large Language Models in Musicology: \\ Are We Ready to Trust the Machines?}

\author{
    Pedro Ramoneda$^{1}$\thanks{ \hspace{0.1cm} Corresponding author: pedro.ramoneda@upf.edu} \hspace{0.2cm}
    Emilia Parada-Cabaleiro$^2$ \hspace{0.2cm}
    Benno Weck$^1$ \hspace{0.2cm}
    Xavier Serra$^1$ \\
    $^1$Music Technology Group, Universitat Pompeu Fabra, Barcelona, Spain \\
    $^2$Department of Music Pedagogy, Nuremberg University of Music, Germany \\
}

\date{}

\begin{document}
\vspace{-2cm}
\maketitle
\vspace{-5cm}

\begin{abstract}
In this work, we explore the use and reliability of Large Language Models (LLMs) in musicology. 
From a discussion with experts and students, we assess
the current acceptance and concerns regarding this, nowadays ubiquitous, technology.
We aim to go one step further, proposing a semi-automatic method to create an initial benchmark using retrieval-augmented generation models and multiple-choice question generation, validated by human experts.
Our evaluation on 400 human-validated questions shows that current vanilla LLMs are less reliable than retrieval augmented generation from music dictionaries. 
This paper suggests that the potential of LLMs in musicology requires musicology driven research that can specialized LLMs by including accurate and reliable domain knowledge.

\end{abstract}

\section{Introduction}

In recent years, 
 research
on Large Language Models (LLMs) has led to 
notable advancements within the text generation domain~\cite{tmlr2023emergent,minaee2024large}. This is the result of training large models on vast non-domain-specific data 
~\cite{gao2020pile,hoffmann2022training}. Well-known families of models include Llama~\cite{llama3modelcard} or GPT~\cite{achiam2023gpt}, which can generate coherent and 
contextually relevant text, making them valuable tools in numerous applications
and professions such as healthcare~\cite{thirunavukarasu2023large}, journalism~\cite{petridis2023anglekindling}, customer support~\cite{kolasani2023optimizing} or education~\cite{kasneci2023chatgpt}. 

Despite their potential, 
LLMs' so-called
hallucinations~\cite{alkaissi2023artificial}, \ie the lack of confidence and accuracy in the text they generate, prevents the use of this technology
in most arts and humanities research tasks~\cite{rane2023role,lozic2023fluent,rane2024role}. 
Issues  include a lack of contextual understanding, bias perpetuation~\cite{gallegos2024bias}, and ethical concerns such as generating misleading content~\cite{weidinger2021ethical}.
The lack of credible source attribution~\cite{rashkin2023measuring} almost render them nugatory for fields like literature, history~\cite{walters2023fabrication}, and law~\cite{weiser2023}.
However, LLMs can aid research through a variety of tasks, such as, translation, text analysis, data organization, historical context retrieval, or summarization.
In this regard,
interdisciplinary research involving the use and further development of LLMs within  
the humanities should be carried out.
This will enable to constructively address existing risks and concerns while developing 
LLMs' full potential, by this delivering their benefits across disciplines.




\begin{figure}[t!]
    \centering
    \resizebox{0.4\textwidth}{!}{ 
    \begin{tikzpicture}
        \node[draw, rectangle, minimum width=8cm, minimum height=1cm, align=center] at (0,1.4) {
            \begin{tabular}{c}
                \textbf{Musicologist:} What's the historical\\ context of this music piece?
            \end{tabular}
        };

        \node[draw, rectangle, minimum width=8cm, minimum height=1cm, align=center] at (0,0) {
            \begin{tabular}{c}
                \textbf{LLM:} It's by Beethoven in 2025! Aliens helped! 
            \end{tabular}
        };

        \node[draw, rectangle, minimum width=8cm, minimum height=1cm, align=center] at (0,-1.4) {
            \begin{tabular}{c}
                \textbf{Musicologist:} I'm not using THIS anymore. 
            \end{tabular}
        };
    \end{tikzpicture}
    }
    \caption{Fictitious interaction illustrating why LLMs' hallucinations might prevent musicologists' trust.}
\vspace{-0.6cm}
\end{figure}

In this work, we focus on musicology,  
a field where the impact of LLMs still needs to be explored.
Musicology, the scholarly study of music, spans from historical research to theoretical analysis~\cite{harap1937nature,groveMusicology}.
Our research mainly focuses on the former,
an area 
which might be greatly supported by LLMs, \eg by breaking language barriers, enhancing information retrieval, or supporting teaching and learning.
However, reliable  sources, such as
music-specialized lexica, monographies, and research articles, 
are often, unlike in  more technical disciplines, not open-access, which prevents LLMs to access high quality information. 
This knowledge deprivation further increases the risk of LLMs to hallucinate, which often leads to non-reliable text generation in musicology related topics.

Through a pilot-survey involving experts and students from the field of musicology, we gather initial insights into the acceptance and trustworthiness of 
LLMs  
in domain-related tasks, 
and its potential impact for music professionals. 
Subsequently, we propose a methodology to measure to which extent  such models posses domain expertise in the field of musicology, by this assessing their practical value for the discipline.
We adopt a Multiple-Choice Question Generation~\cite{liu2024chatqa} approach to semi-automatically construct a benchmark leveraging recent advancements in retrieval-augmented generation models~\cite{lewis2020retrieval}.
To automatically generate 
high-quality questions, we provide the generation model with domain-knowledge from \textit{The New Grove Dictionary of Music and Musicians}~\cite{grove2001music}, 
an established and reliable source. 
The final benchmark, made up by 400  question-answer pairs  validated by a human expert, 
is evaluated on several open-source models.  
This dual approach---survey and benchmark---provides a comprehensive understanding of the challenges and potential solutions for meaningful integration of LLMs in musicology.

\section{Pilot-survey:  LLMs in musicology}

We conducted a survey targeting professionals related to musicology. The survey included questions to identify the respondent's domain of study (e.g., musicology, composition, music pedagogy, music performance), the highest level of music education completed or being pursued, and their familiarity with technologies known as LLMs such as ChatGPT. Additionally, the survey inquired about the frequency of interactions with LLMs, particularly in the context of musical topics like Music Theory and Music History. Participants were asked to rate the trustworthiness and usefulness of LLMs for these subjects, as well as to consider its 
revolutionary impact 
on the field of musicology. 
Lastly, the survey explored the 
possible consequences of LLMs on music professionals, both presently and in the future.

\begin{figure}
    \centering
    \includegraphics[width=1.0\linewidth]{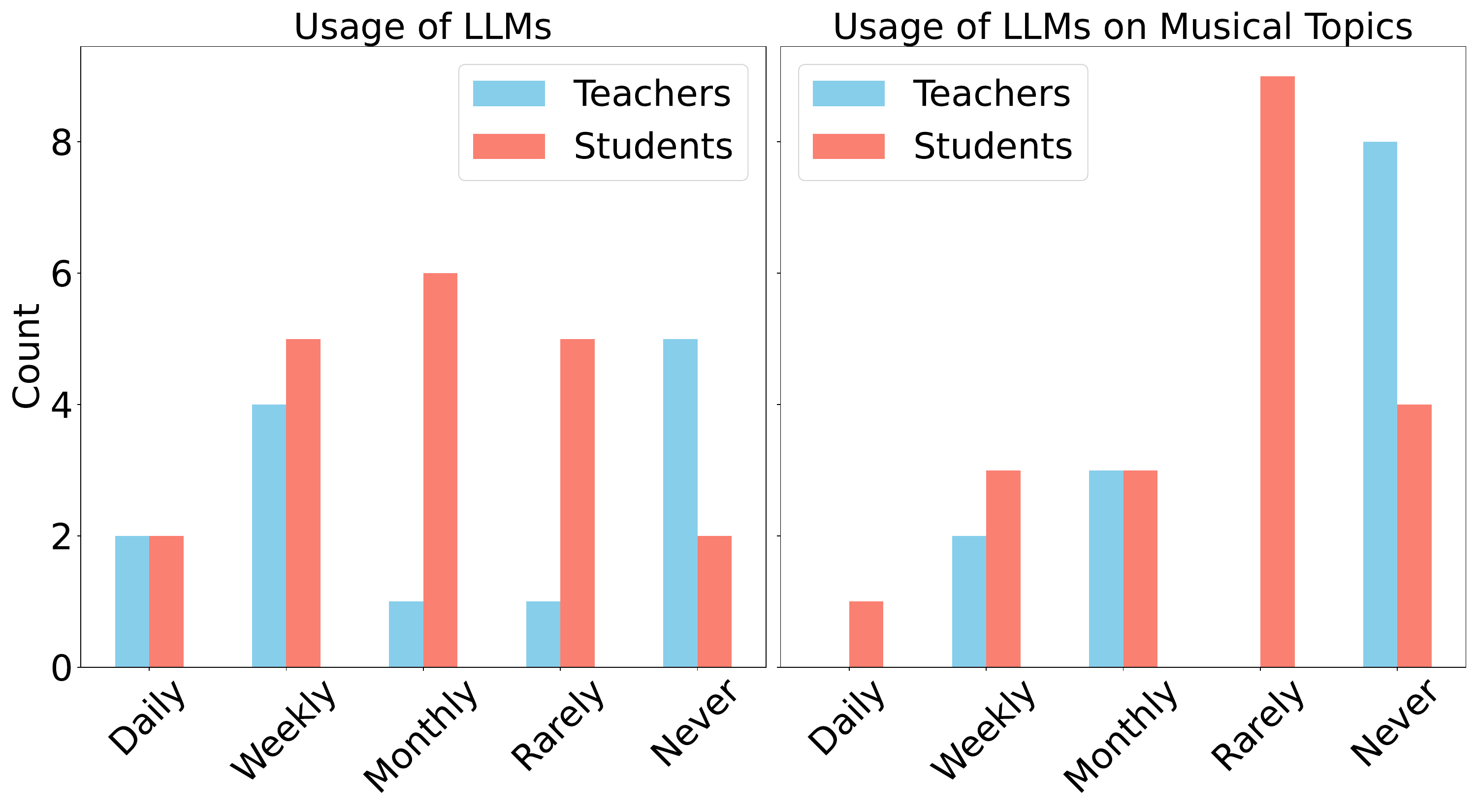}
    \vspace{-0.8cm}
    \caption{Survey's answers about the usage of LLMs in general (left) and  on music topics (right)}
    \label{fig:interaction}
\vspace{-0.3cm}
\end{figure}

A total of 33 participants, having or pursuing a Bachelor's degree in music, completed the survey: 
20 students, 7 lecturers, 11 researchers, and 8 music educators (multiple areas can be selected). 
In terms of discipline, the respondents are distributed as the following: 
22 Musicology and Related Studies, 10 Music Performance, 3 Music Pedagogy, 2 Composition, 1 Conduction, and 1 Music Therapy. 
While only one participant (from the field of musicology) had not heard about LLMs before, in terms of the  participants' frequency of use 
and trustworthiness, 
a noticeable gap between students and teachers\footnote{For simplicity, with `teachers' we refer to all the participants who did not identified themselves as student.} can be observed. 

\begin{figure}
    \vspace{-0.25cm}
    \centering
    \includegraphics[width=1.\linewidth]{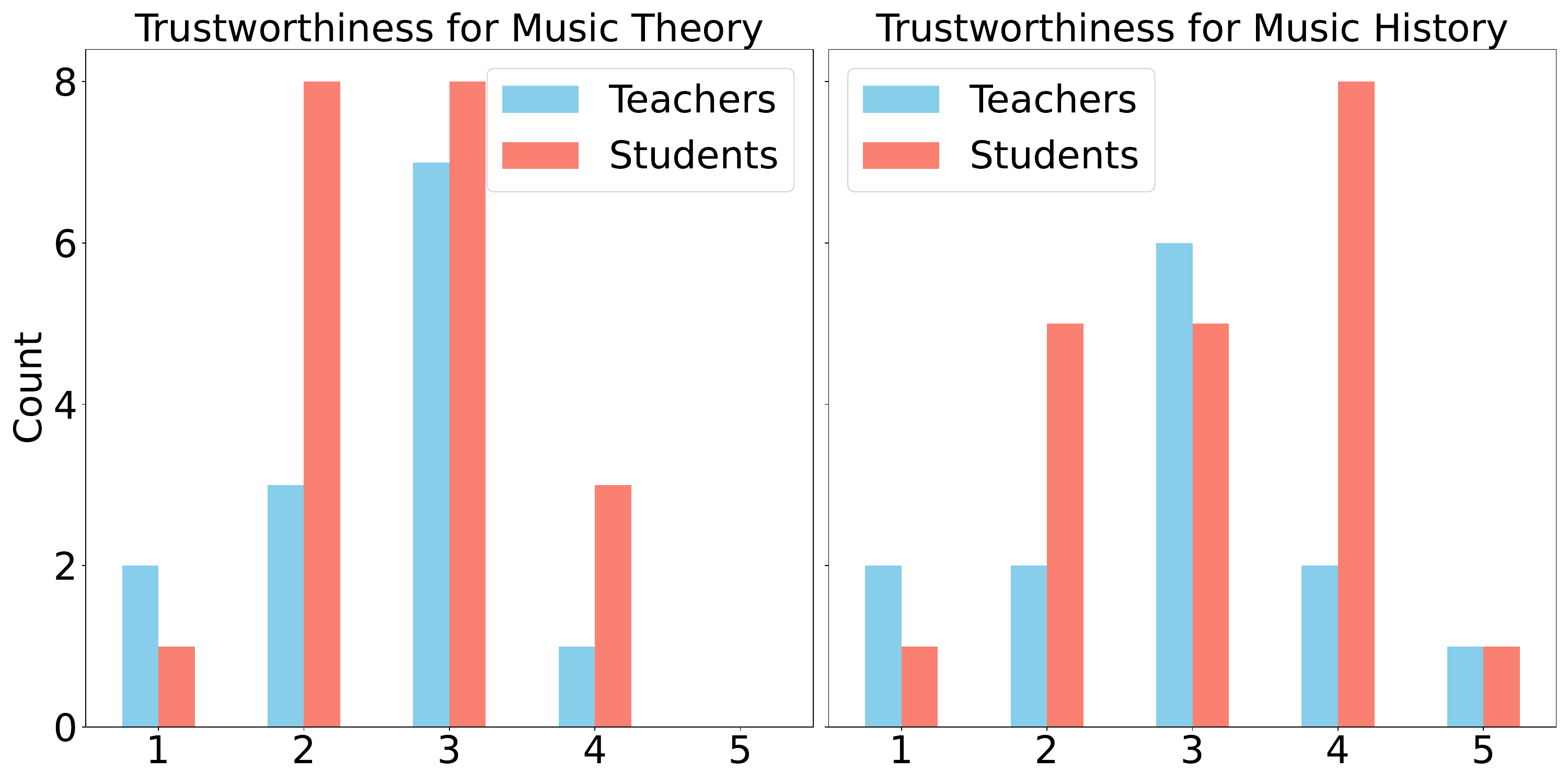}
    \vspace{-0.7cm}
    \caption{Survey's answers about the usage of LLMs in general (left) and  on music topics (right).}
    \label{fig:trust}
\end{figure}

Figure~\ref{fig:interaction} 
illustrates how often students and teachers interact with LLMs in general and about music topics.
Teachers 
frequently or not at all, while most students use it weekly or monthly.
However, both groups tend to not use LLMs for music-related topics. 
Participants' judgement of LLMs' trustworthiness is depicted in Figure~\ref{fig:trust} and the trend of ratings is similar across both groups.
Additionally, confidence in LLMs is slightly higher for Music History than for Music Theory, indicating a nuanced perception of their reliability in different musicology subfields.

\begin{figure}
    \centering
    \includegraphics[trim={2.9cm  1.5cm 0cm 0cm},clip,width=\columnwidth]
    {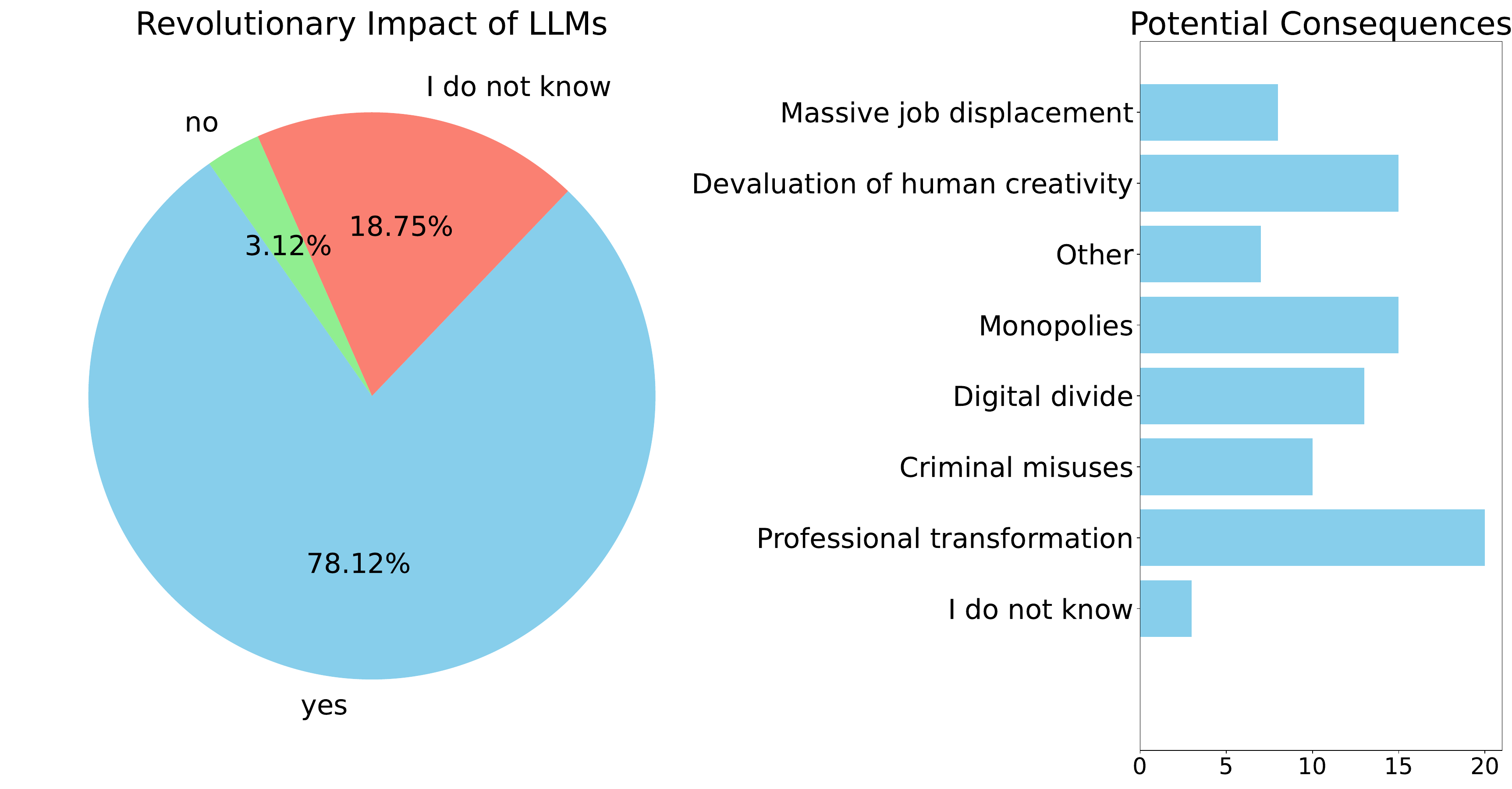}
    \caption{Survey answers about the revolutionary impact (right) and potential consequences (left) of LLMs.}
    \label{fig:potential}
\vspace{-0.3cm}
\end{figure}

Most of the participants
(78\%)
agreed that LLMs might revolutionize the field of musicology (cf.\ Figure~\ref{fig:potential}, left).
While the anticipated potential consequences of LLMs for the field are varied,
professional transformation seems to be
the most prominent (20 votes), as illustrated in the histogram 
(cf.\ Figure~\ref{fig:potential}, right).
In conclusion, despite limited current usage and trust, experts anticipate a significant future impact of LLMs on musicology, motivating current research on the topic. 

\section{Musicology Benchmark: TrustMus}

This section outlines our strategy for evaluating how much LLMs 
hallucinate in musicology. It summarizes the creation of the human-validated multiple-choice benchmark \textit{TrustMus}, \ie a  collection of reliable questions related to various musical topics and concepts, and analyzes the models' performance on the benchmark. 



Following previous works~\cite{halueval}, multiple-choice questions are generated after  extracting relevant information from a text source: here, 
\textit{The Grove Dictionary Online} \cite{grove2001music}.\footnote{The Grove Dictionary is a copyrighted work. Using its content for generating questions is under fair use for research purposes. The EU Directive on Copyright in the Digital Single Market allows text and data mining for research purposes.} 
In order to identify the most relevant articles within the text source, we used a PageRank-like algorithm~\cite{hagberg2008exploring}. 

\begin{figure}
    \vspace{-0.25cm}
    \centering
    \includegraphics[width=0.85\linewidth]{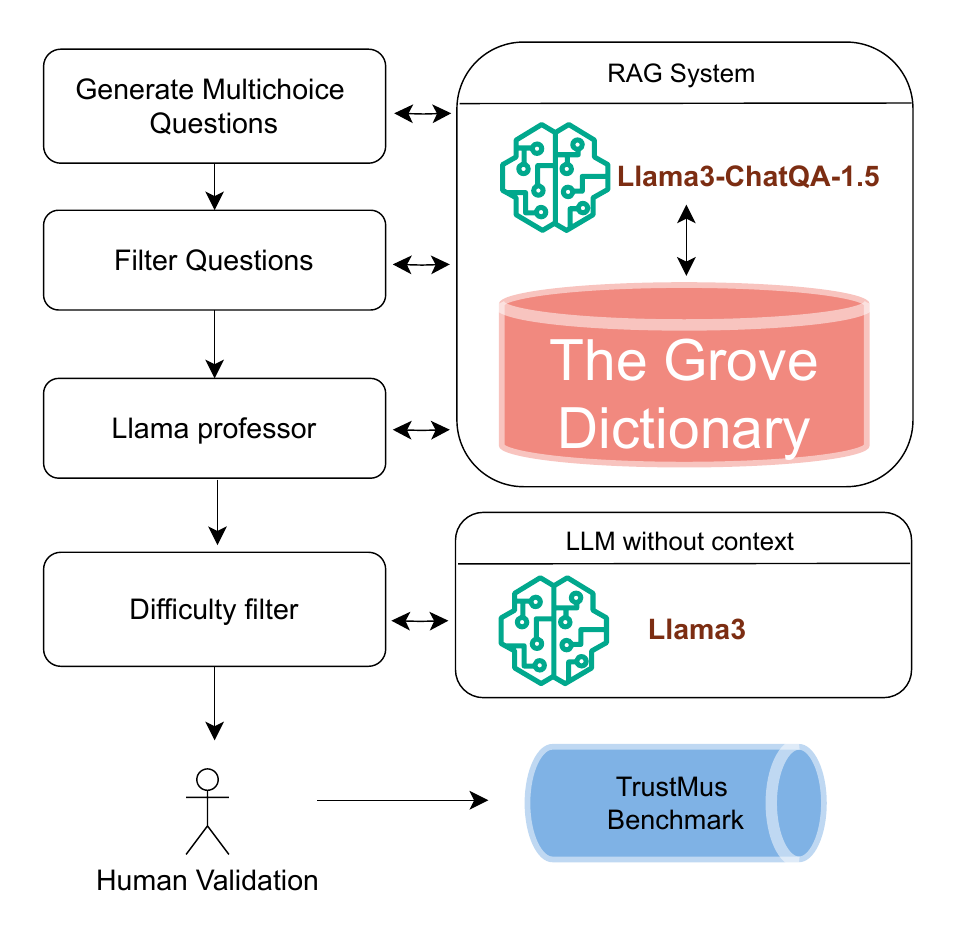}
    \vspace{-0.4cm}
    \caption{Language chain for generating the multi-choice questions.}
    \label{fig:chain}
    \vspace{-0.3cm}
\end{figure}

To accelerate the creation of TrustMus, we designed a workflow inspired by recent works~\cite{correctiverag,adaptiverag,asai2023selfrag,dhuliawala2023chain}, as shown in Figure \ref{fig:chain}.
First, we generated five questions from each article, each with four possible answer options, using a fine-tuned LLM for retrieval-augmented generation (RAG) \cite{liu2024chatqa}, resulting in 7\,500 questions.
Second, we discard questions that did not have relation with musicology or a unique and unambiguous answer, by prompting the same LLM to decide based on the article, eliminating 2\,632 questions.
Next, we attempted to answer the remaining questions using a RAG-like model that we term \textit{Llama Professor} by giving the article as context to the LLM.
Questions for which Llama Professor chooses the wrong answer option are considered ambiguous or unusable and are thus removed, resulting in 3\,285 valid questions.
All previous prompts used the Chain of Thought (CoT) method to enhance the model's reasoning skills \cite{cot}.
Before human intervention, we attempted to answer the questions with llama3-8B~\cite{llama3modelcard} without RAG and in one shot, \ie without the chain of thought (cf.\ the difficulty filter  in Figure~\ref{fig:chain}), which lead  an accuracy of 67.4\%.
Thus, arguably simple questions are eliminated, resulting in 1\,081 domain-ones.

The resulting set of questions 
was automatically classified   with a CoT prompt into four classes, according to their topic:  People (Ppl);  Instruments and Technology (I\&T);  Genres, Forms, and Theory  (Thr);  Culture and history  (C\&H).
An expert human 
annotator validated questions until  100 valid ones per class were identified (on average, 17\% of those assessed were discarded).\footnote{@: \url{https://zenodo.org/records/13644330}}

\vspace{-0.2cm}
\section{Results and Discussion}

\subsection{Human validation insights}

\begin{table*}[ht!]
\vspace{-0.5cm}
\centering
\small
\setlength{\tabcolsep}{4pt} 
\begin{tabular}{cccccccccc}
\toprule
\textbf{Model} & \textbf{Quant} & \textbf{TrustMus} & \textbf{Rank}&\textbf{Ppl} & \textbf{I\&T} & \textbf{Thr} & \textbf{C\&H} & \textbf{LB} & \textbf{Rank}\\
\cmidrule(r){1-8} \cmidrule(lr){9-10}
gpt-4o-2024-05-13~\cite{achiam2023gpt}  & API & 58.75 & 1 &  60.0 & 44.0 & 61.0 & 70.0 & 58.38 & 1  \\ 
mixtral-8x7b-instruct-v0.1~\cite{jiang2024mixtral} & \ding{51} &  40.5\phantom{0} &  2 &41.0 & 30.0 & 43.0 & 48.0 & 37.24 & 4 \\ 
gpt-3.5-turbo-0125~\cite{achiam2023gpt} & API & 39.75 & 3 & 39.0 & 25.0 & 43.0 & 52.0 & 37.97 & 5 \\ 
meta-llama-3-70b-instruct~\cite{llama3modelcard} & \ding{51} & 37.75 & 4 & 41.0 & 23.0 & 44.0 & 43.0 & 42.76 & 2 \\ 
qwen2-72b-instruct~\cite{bai2023qwen} & \ding{51} & 35.5\phantom{0} & 5 & 39.0 & 27.0 & 37.0 & 39.0 & 41.43 & 3 \\ 
qwen2-7b-instruct~\cite{bai2023qwen} & \ding{55} & 34.0\phantom{0} & 6 & 29.0 & 43.0 & 36.0 & 41.0 & 25.92 & 9 \\ 
phi-3-medium-4k-instruct~\cite{abdin2024phi}  & \ding{51} & 32.75 & 7 & 32.0 & 27.0 & 38.0 & 34.0 & 33.46 & 6 \\ 
meta-llama-3-8b-instruct~\cite{llama3modelcard} & \ding{55} & 32.75 & 8 & 43.0 & 22.0 & 31.0 & 35.0 & 31.05 & 7 \\ 
phi-3-small-128k-instruct~\cite{abdin2024phi}  & \ding{55} & 31.5\phantom{0} & 9 & 20.0 & 29.0 & 41.0 & 36.0 & 28.12 & 8 \\ 
\addlinespace
Llama Professor (RAG)  &\ding{55} & 100.0 &  - & 100.0 & 100.0 & 100.0 & 100.0 & - & - \\
\bottomrule
\end{tabular}

\caption{Benchmark results (accuracy) on the 400 validated 
questions (TrustMus) and per  category:  People (Ppl);  Instruments and Technology (I\&T);  Genres, Forms, and Theory (Thr);  Culture and History (C\&H). Whether the models are quantized (Quant), their rank, and  LiveBench
average score (LB) excluding math ranking is also given.}
\label{tab:model_comparison}
\vspace{-0.5cm}
\end{table*}

Some examples of hallucinations of Llama3 without RAG and CoT, the difficulty filter, are as follows: \textit{What does the natural sign ($\natural$) do in music notation? A) Raises a note by one semitone, B) Raises a note by two semitones, C) Lowers a note by one semitone, D) Cancels a previous sharp or flat.}  The correct answer is D, but Llama3 chose A, which any musician should know is incorrect. 

Another type of limitation of LLMs in the context of musicology, is the need of the models for interpreting the information.
This can be illustrated by how Llama3 handled the
article about \textit{Adagio} in the \textit{Grove Dictionary Online}, 
which summarizes the evolution of the term over centuries. %
In this regard, when interrogating Llama3 about the term as described by Rousseau, the model refers to the modern definition.

\subsection{TrustMus evaluation}

Table~\ref{tab:model_comparison} presents the benchmark results for various models evaluated on TrustMus.\footnote{Since we believe that open models are critical for transparency, reproducibility, and the advancement of knowledge, we use them in  our research. We included ChatGPT in our comparison only because it is currently the most used LLM.} 
The models tested include the best open source performing models in 
LiveBench -- LB~\cite{livebench} excluding coding and math categories, \ie a benchmark for LLMs without contamination and reduced biases containing  non-musicology knowledge. Due to its' leading performance, 
results of OpenAI's GPT models are also given for comparison.
Models with less than 8B parameters were deployed in a computer with two RTX 2080ti GPUs with 16-bit precision, the largest models in a Colab A100 GPU with 4-bit quantization, and the 
GPT models through their official API.

The model \texttt{gpt-4o-2024-05-13} clearly outperforms others with an accuracy of 58.75\% (cf.\ TrustMus score in Table~\ref{tab:model_comparison}),  excelling in the categories 
Ppl, Thr, and C\&H.  This is not surprising as it is the leading model in LB  as well, 
with a score of 58.38\%. 
However,  comparing the LB and TrustMus rankings   reveals important differences about how the models perform in terms of  general and in domain-specific knowledge. 
For instance, unlike in LB, the model \texttt{mixtral-8x22b-instruct-v0.1} performs well in our benchmark, ranking second with a score of  40.5\%.
It is important to note the similar performance between \texttt{qwen} with 72B and 7B, the latter  being the best performing of the `small' LLMs in TrustMus while showing the worst performance in LB. 
We also aim to acknowledge that comparing \texttt{meta-llama3}
models with others is not entirely fair, as the benchmark was automatically generated by selecting questions from their specific blind spots, as detailed in Section 3.


The lower performance of open-source models compared to LLama professor (RAG)~\cite{liu2024chatqa} highlights
the importance of reliable
domain-specific knowledge
for musicology-related applications. 
This indicates considerable improvement possibilities with the potential of increasing the trustworthiness of LLMs 
in the field. 


\vspace{-0.2cm}
\section{Conclusions}
\vspace{-0.2cm}


Our paper shows that while current usage and trust in LLMs in musicology are low, there is a strong expectation of future impact.
However, LLMs are not yet at the required level for the field 
and do not meet the minimum quality, ethical and likely legal standards currently being discussed. 
Through the proposed semi-automatic benchmark, we present a first attempt to  measure LLMs  hallucinations on musicology-related tasks.
This approach aims to facilitate the evaluation of future models, which promotes transparency and trustworthiness of the technology.  
%
Despite the effort, this initial experiments 
are insufficient. Besides 
a more thorough evaluation, 
there is the need to 
specialize current models for musicology-related tasks, while 
reducing their environmental footprint.
%
Further research should focus on ensuring LLMs reliability to avoid misinformation, protecting user privacy and data security, and mitigating training data biases to promote responsible use in musicology. Collaboration between the technological, musicological, and content owner communities is essential for the proper development of this technology.






\bibliography{nlp4MusA}
\bibliographystyle{nlp4MusA_natbib}

\end{document}